\documentclass[journal=jctcce]{achemso} 
\usepackage[latin1]{inputenc}
\usepackage[english]{babel}
\usepackage{amsmath, amsfonts, amssymb, graphicx, subfigure, xcolor}
\usepackage[labelfont=bf]{caption} 
\usepackage{subfig}

\newcommand{\added}{\textcolor{black}}

\usepackage[skip=4pt, labelfont=bf]{caption}

\newcommand{\beginsupplement}{%
        \setcounter{table}{0}
        \renewcommand{\thetable}{S\arabic{table}}%
        \setcounter{figure}{0}
        \renewcommand{\thefigure}{S\arabic{figure}}%
     }

\author{Hao Tian}
\affiliation{Department of Chemistry, Center for Scientific Computation, Center for Drug Discovery, Design, and Delivery (CD4), Southern Methodist University, Dallas, Texas, United States of America}

\author{Peng Tao}
\affiliation{Department of Chemistry, Center for Scientific Computation, Center for Drug Discovery, Design, and Delivery (CD4), Southern Methodist University, Dallas, Texas, United States of America}
\email{ptao@smu.edu}

\title{Deciphering the Protein Motion of S1 Subunit in SARS-CoV-2 Spike Glycoprotein Through Integrated Computational Methods}

\begin{document}

\maketitle


\begin{abstract}

The novel severe acute respiratory syndrome coronavirus 2 (SARS-CoV-2) is a major worldwide public health emergency that has infected over $1.5$ million people. The partially open state of S1 subunit in spike glycoprotein is considered vital for its infection with host cell and is represented as a key target for neutralizing antibodies. However, the mechanism elucidating the transition from the closed state to the partially open state still remains unclear. Here, we applied a combination of Markov state model, transition path theory and random forest to analyze the S1 motion. Our results explored a promising complete conformational movement of receptor-binding domain, from buried, partially open, to detached states. We also numerically confirmed the transition probability between those states. Based on the asymmetry in both the dynamics behavior and backbone C$\alpha$ importance, we further suggested a relation between chains in the trimer spike protein, which may help in the vaccine design and antibody neutralization. 

\end{abstract}

\section{Introduction}

As of April 9 2020, there has been over $1.5$ million confirmed cases of the newly discovered coronavirus, named as SARS-CoV-2, causing over $90,000$ cases of death, according to the World Heath Organization (WHO). SARS-CoV-2 is related to bats-derived coronaviruses and the SARS-CoV, reported in the Guangdong province of China in $2002$, and identified as a new member of betacoronavirus. \cite{wu2020new,lu2020genomic} Due to its fast spreading though human contacts, the WHO declared it as a Public Health Emergency of International Concern (PHEIC). 

SARS-CoV-2 is known to infect human through the interaction between its spike (S) protein and human host receptors. \cite{cavanagh1995coronavirus,lu2015bat,wang2016mers} The S protein is a trimer (chain A, B and C) and each chain is formed by S1 and S2 subunits that are related with host receptors binding and membranes fusion, respectively. \cite{walls2020structure,li2016structure,li2015receptor} The S1 subunit consists of a N-terminal domain (NTD), receptor-binding domain (RBD) and two subdomains (SD1 and SD2). \cite{wrapp2020cryo} It is reported that RBD undergoes a conformational change from stable close state to dynamically-less-favorable partially open state in chain A. \cite{li2016structure,bosch2003coronavirus} In the closed state, the determinants of receptor binding are buried and inaccessible to receptors, while in the partially open state are exposed and is expected to be necessary for the interaction with host cells. \cite{gui2017cryo,pallesen2017immunogenicity} In the cases of SARS-CoV-2 and SARS-CoV, S glycoprotein is found to inherently sample closed and open states and this behavior is suggested to exist in the most pathogenic coronaviruses. \cite{walls2020structure,shang2018cryo} \added{While the partially open state plays an important role in human cell infection, little study is done illustrating this protein motion on residue level, which could potentailly facilitate the prevention and treatment.}

Molecular dynamics (MD) simulations can provide atomic scale information and are widely used in sampling protein movement and structure landscape. \cite{prinz2011markov} Two kinds of trajectories from closed and open states of SARS-CoV-2 S protein initiating from close state (PDB 6VXX) and partially open state (PDB 6VYB) are available from D E Shaw Research \cite{DEShaw}. However, the timescale ($10$ microseconds) is still relatively trivial compared with the timescale of biological processes in the real world. To gain more information from the result of MD simulation, Markov state model (MSM) is applied to catch long-time kinetic information given time-limited simulation trajectories. \cite{suarez2016accurate,adelman2016stochastic} Another advantage of MSM is that it can divide a large number of protein structures from simulation into subspaces based on the extracted kinetic information, and differences among those spaces can be calculated and used for comparison.

Machine learning techniques have achieved great accomplishments in chemistry and biology including material discovery, \cite{raccuglia2016machine} structure representation \cite{faber2015crystal} and computation acceleration \cite{botu2015adaptive}.  The great contributions from machine learning mainly come from its ability to deal with large scale data and its accurate and explainable models, \cite{kotsiantis2007supervised,jing2018deep} which provide a new opportunity to decipher protein. In this study, tree-based machine learning models were used to identify important residues. Specifically, random forest model was applied as a classification model to classify different structures and calculate the contribution of each residue and structure importance for the close-open transition process. 

The transition from close state to partially open state of S1 subunits of SARS-CoV-2 S protein is investigated in this research though a series of Markov state model, transition path theory and random forest. Our results numerically confirmed the close-open transition probability through estimated transition matrix, showed a complete transition path from close to open state and identified a relationship between the motion of chain A and two other chains.

\section{Methods}

\subsection{Analysis of Simulation Trajectories}

The root-mean-square deviation (RMSD) is used to measure the overall conformational change of each frame with regard to a reference structure in a MD simulation trajectory. For a molecular structure represented by Cartesian coordinate vector $r_i$ ($i = 1 \text{  to  } N$) of $N$ atoms, the RMSD is calculated as following: 

\begin{equation}
\text{RMSD} = \sqrt{\frac{\sum_{i=1}^N (r_i^0 - Ur_i)^2}{N}}
\end{equation}

$r_i^0$ represents the Cartesian coordinate vector of the $i^{\text{th}}$ atom in the reference structure. The transformation matrix $U$ is defined as the best fit alignment between the protein structures along trajectories with respect to the reference structure. 

The root-mean-square fluctuation (RMSF) is used to measure the fluctuation of each atom in each frame with regard to a reference structure in a MD simulation trajectory. 

\begin{equation}
\text{RMSF}_i = \sqrt{\frac{1}{T} \sum_{j=1}^{T} (v_i^j - \bar{v_i})^2}
\end{equation}

Where $T$ is the total frames and $r_i$ is the average position of atom $i$ in the given trajectory. 

\subsection{Feature Processing}

Distances between pairs of backbone C$\alpha$ were chosen as representative features for one structure. Specifically, each distance pair of one C$\alpha$ carbon and all other C$\alpha$ carbon in all amino acids was calculated. A protein contact map technique is normally built by transforming each pair-wised distance value into $1$ if that feature value is below a threshold or $0$ otherwise \cite{doerr2017dimensionality}. However, this feature preprocessing technique risks to ignore potentially useful spacial information forcing a boolean value on the features. Therefore, inspired by ReLU activation function \cite{nair2010rectified} in neural network, \added{which the equation is shown below}, we proposed a revised feature transformation method by transforming each feature value into 0 if that feature is above a threshold and keep it the same value otherwise. Compared with reference feature transformation rule, our proposed technique can still build a protein contact map while can differentiate local features with the least local information loss.

\begin{equation}
f(z) = \max{(0, z)}
\end{equation}

\subsection{Random Forest Model}

Random forest is a machine learning technique that can be used for classification. \cite{liaw2002classification,wang2019machine} A random forest is composed of several decision trees, \cite{utgoff1989incremental} which are trained based on given training data. The final classification output of a random forest model is a collection of classes predicted by each decision tree model. The random forest algorithm carried out in this study is implemented in scikit-learn \cite{pedregosa2011scikit} program package version 0.20.1. The number of decision trees used was 50. One advantage of random forest model over decision tree model is that employing multiple decision tree models mitigates the overfitting problem suffered by single decision tree model. 

\subsection{Feature Importance}

In a random forest model, a quantitative evaluation of the importance for each feature used for training
is calculated through training process. This feature importance is calculated using Gini impurity:

\begin{equation}
\text{Gini impurity} = \sum_{i=1}^C -f_i (1-f_i)
\end{equation}

where $f_i$ is the frequency of a label at a node of being picked to split the data set and $C$ is the total number of unique labels. A random forest model is a collection of several decision tree models. The importance of node $i$ in decision tree is calculated as:

\begin{equation}
n_i = w_i C_i - \sum_i^m w_{m(i)} C_{m(i)}
\end{equation}

where $w_j$ is the weighted number of samples reading node $i$, $C_i$ is the impurity value of node $i$ and $m$ is the number of child nodes. The feature importance of feature $i$ is calculated as

\begin{equation}
f_i = \frac{\sum_l^s n_j}{\sum_{\text{k} \in \text{all nodes}} n_k} 
\end{equation}

where $s$ is the times of node $j$ split on feature $i$. Feature importance within a decision tree is further normalized by:

\begin{equation}
\text{norm} f_i = \frac{f_i}{\sum_{j \in \text{all features}} f_j}
\end{equation}

The feature importance in random forest is the averaged importance of feature $i$ in all decision tree models:

\begin{equation}
F_i = \frac{\sum_{j \in \text{all decision trees}} \text{norm} f_i}{N}
\end{equation}

where norm $f_i$ is the normalized feature importance of one single decision tree and $N$ is the number of decision trees. \cite{breiman2001random}

Feature importance of all pairwise C$\alpha$ distances were calculated using the above methods. The feature importance of an amino acid is the summation of importances of features that are related with that amino acid. The relative accumulated feature importance of each amino acid shows the distinguishability and contribution of that amino acid among all amino acids in differentiating states.

\subsection{Markov State Model}

Markov state model (MSM) is used to construct long-timescale dynamics behavior \cite{wang2019dynamical}. MiniBatch k-means clustering  was used to classify each simulation frame to microstates. Perron-cluster cluster analysis (PCCA) \cite{deuflhard2005robust} was applied to cluster microstates into macrostates. Macrostates are considered as equilibrium or steady states. Transition matrix and transition probability were calculated to quantitatively show the transition dynamics between macrostates. A specific time interval, referred to as lag time, needs to be determined to construct transition matrix. The value of the lag time, as well as the number of macrostates, is selected based on the result of estimated relaxation timescale. \cite{bowman2009using} MSMBuilder \cite{harrigan2017msmbuilder} version 3.8.0 was employed to build markov state models in this study.

\subsection{Transition Path Theory}

Transition path theory (TPT) \cite{noe2009constructing, metzner2009transition} is used to calculate the probability of transitioning from one state to another within the framework of a MSM. \added{In the current study, the most representative structure of closed state was chosen among the $4$ macrostates in the closed state and defined as macrostate $2$ due to the reason that it was the most probable macrostate to transfer to other open macrostates. The open state was chosen based on the stability of each open macrostates and macrostate $8$ was chosen as representative open state.} All other states are treated as intermediate states. Possible transition pathways from the closed to the open state were explored. The committor probability $q_i^+$ is defined as the probability from state $i$ to reach the target state rather than initial state. Based on definition, $q_i^+ = 0$ for all microstates in initial state and $q_i^+ = 1$ for all microstates in target state. The committor probability for all other microstates are calculated by the following equation:

\begin{equation}
-q_i^+ + \sum_{k \in I} T_{ik} q_k^+ = -\sum_{k \in \text{target state}} T_{ik}
\end{equation}

where $T_{ik}$ is the transition probability from state $i$ to state $k$. Sequentially, the effective flux is calculated as:

\begin{equation}
f_{ij} = \pi_i q_i^- T_{ij} q_j^+ 
\end{equation}

where $\pi_i$ is the equilibrium probability of normalized transition matrix  $T_{\pi}$ and $q_i^-$ is the backward-committor probability calculated as $q_i^- = 1 - q_i^+$. However, backward flux $f_{ji}$ should also be considered and subtracted when calculating net flux. Therefore, the net flux $f_{ij}^+ =  max(0,  f_{ij} - f_{ji})$. Total flux can then be calculated as:

\begin{equation}
F = \sum_{i \ \in \text{ initial state }} \sum_{j \ \notin \text{ initial  state }} \pi_i T_{ij} q_j^+
\end{equation}

The flux from initial state to target state can be decomposed to individual pathways $p_i$. Dijkstra algorithm is implemented in MSMBuilder for pathway decomposition. A set of pathways $p_i$ can be generated along $f_i$, which provides a relative probability by:

\begin{equation}
p_i = \frac{f_i}{\sum_j f_j}
\end{equation}

\section{Results}

\subsection*{Simulation trajectory analysis shows dynamical activity}

Two $10$ microseconds simulation trajectories of the trimeric SARS-CoV-2 S glycoprotein were treated as reference and backbone alpha carbons (C$\alpha$) of the trimer were chosen and extracted as representative features of structures.

\begin{figure*}[t]
	\begin{minipage}[c][\width]{0.49\textwidth}
	   \centering
	   \includegraphics[width=\textwidth]{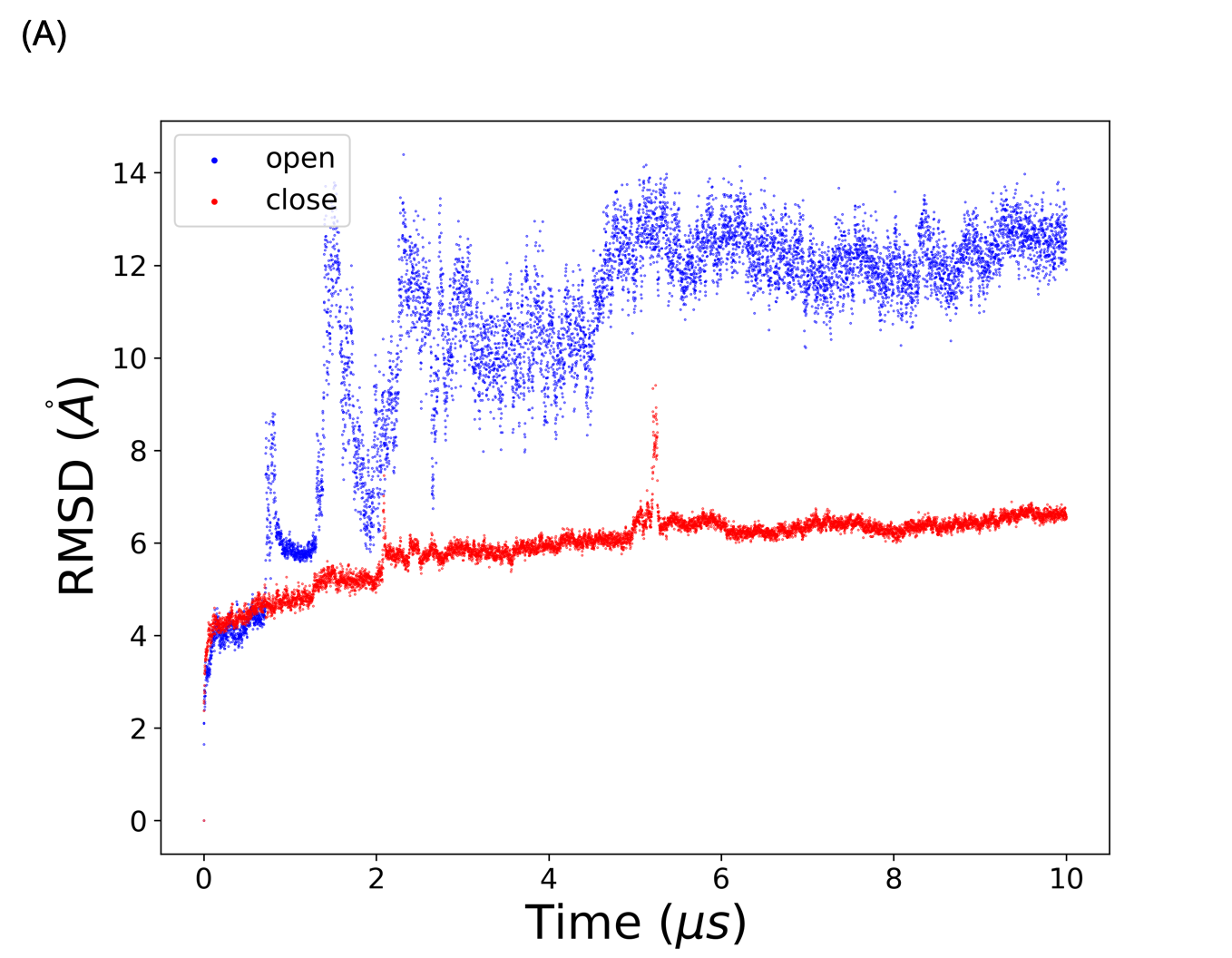}
	   \label{RMSD}
	\end{minipage}
 	\hfill 	
	\begin{minipage}[c][\width]{0.49\textwidth}
	   \centering
	   \includegraphics[width=\textwidth]{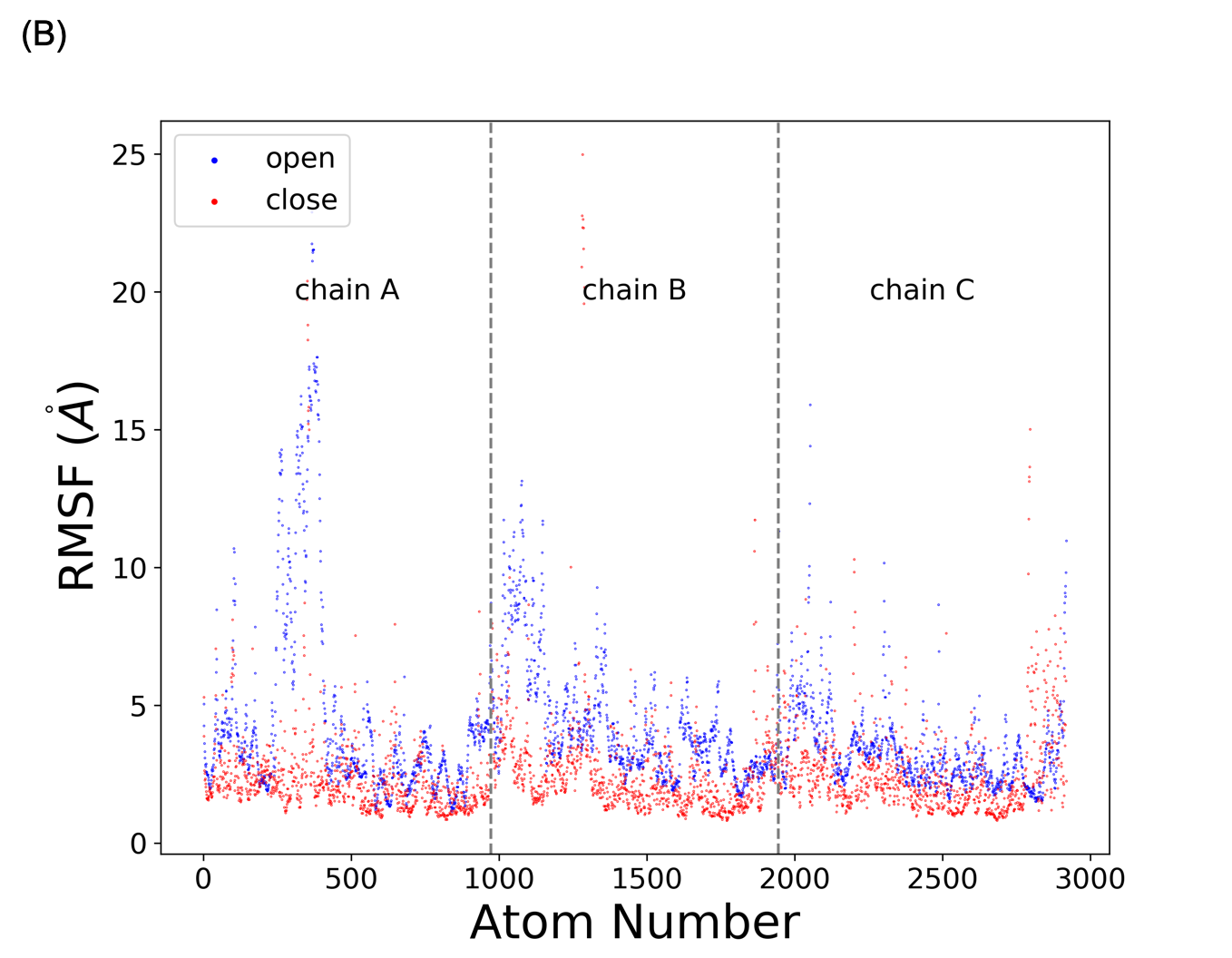}
	   \label{RMSF}
	\end{minipage}
\vspace{-2em}
\caption{Simulations of SARS-CoV-2 S glycoprotein: (A) RMSD of the trajectories in the close (red) and open (blue) states. (B) RMSF of simulation trajectories in the close (red) and open (blue) states. The protein is divided into three chains separated by grey dashed lines. Atom number was counted from 0.}
\label{RMSD_RMSF}
\end{figure*}

To probe the dynamical stability of two structures, the time evolution of the RMSD were plotted in Figure \ref{RMSD_RMSF} (A). All RMSD values were calculated with reference to the first frame of each trajectory. The average RMSD values in two states are $5.9\mathring{A}$ and  $10.6\mathring{A}$, respectively. The plot suggested that the close state is relatively stable while the partially open state is dynamically active and undergoes significant conformational changes after $1$ microsecond. However, the simulation of open state after 6 microseconds suggested a convergence in the RMSD value and a relatively stable structure, which corresponds to the detached S1 subunit from S2 fusion machinery. 

RMSF results were plotted in Figure \ref{RMSD_RMSF} (B). The asymmetry in protein motion was noticed by comparing the individual dynamics behavior among three chains. Corresponding to the RMSD result in chain A, the RMSF result showed a similar high-degree conformational change in the RBD domain. However, through the detachment in chain A, chain B and C showed different movements that both NTD and RBD in chain B are more dynamically active than those in chain C. while in closed state the chains B and C displayed similar dynamics.

\subsection*{Markov state model and transition path theory elucidates the close-open transition}

\begin{figure*}[t]
	\begin{minipage}[c][\width]{0.49\textwidth}
	   \centering
	   \includegraphics[width=\textwidth]{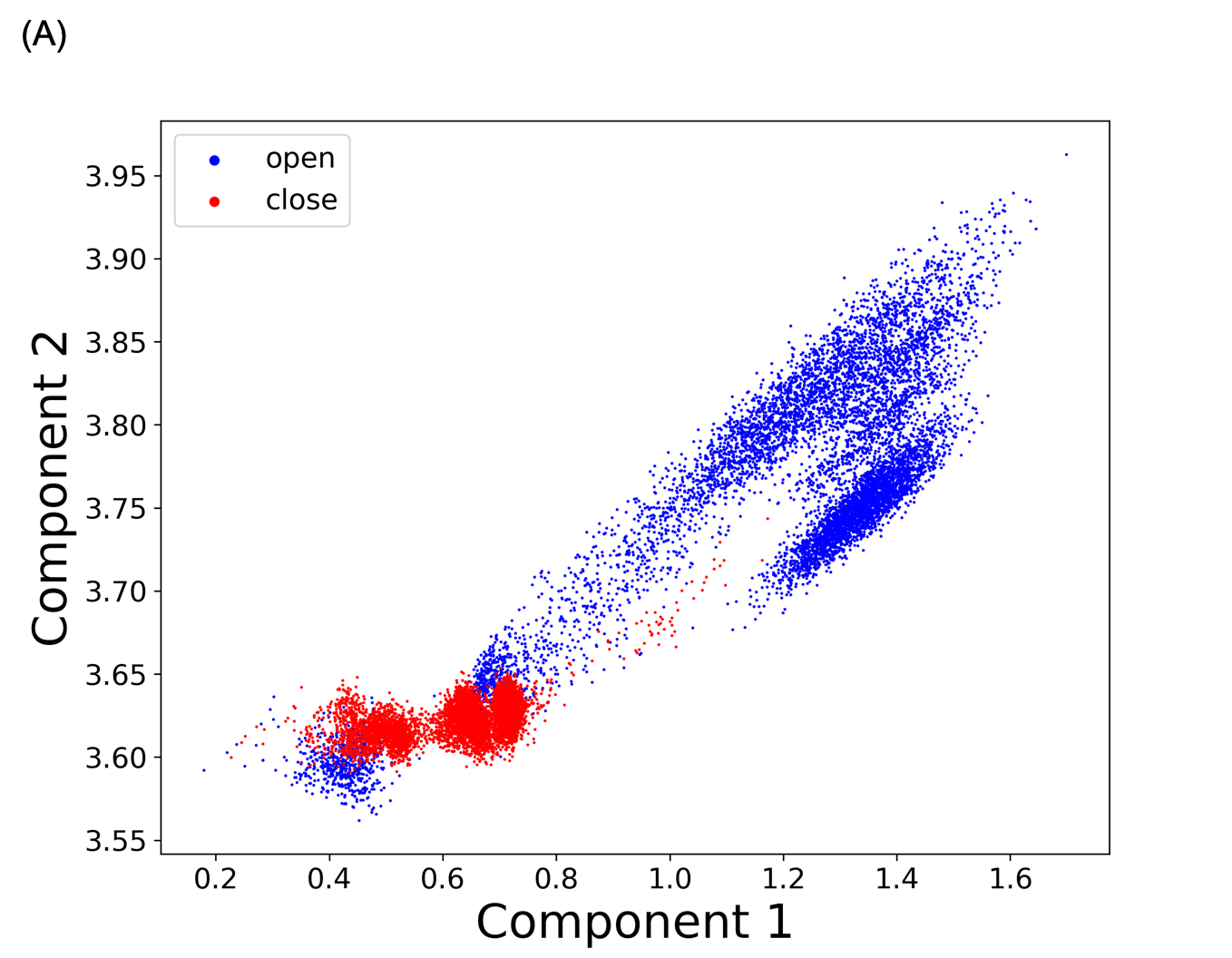}
	\end{minipage}
 	\hfill 	
	\begin{minipage}[c][\width]{0.49\textwidth}
	   \centering
	   \includegraphics[width=\textwidth]{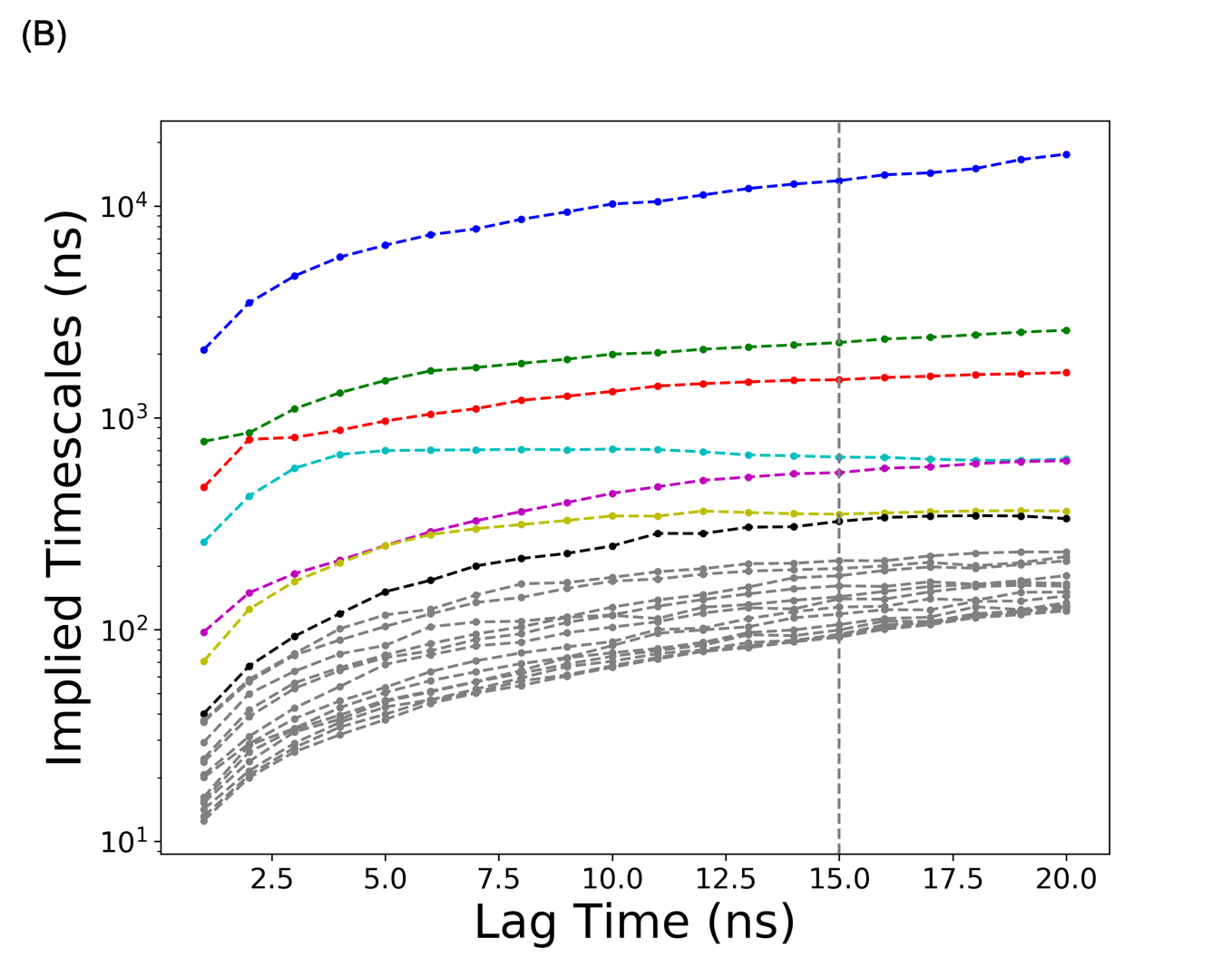}
	\end{minipage}
\vspace{-2em}
\caption{Distribution of SARS-CoV-2 S glycoprotein simulations: (A) 2D protein conformation space using RMSD values with references both closed and partially open states. (B) Implied relaxation timescale of top $20$ variables based on the data coordinates in the reduced map regarding with the different lag time as interval.}
\label{reduced2d_timescale}
\end{figure*}

\begin{figure}[ht]
	\centering
	\includegraphics[width=0.9\textwidth]{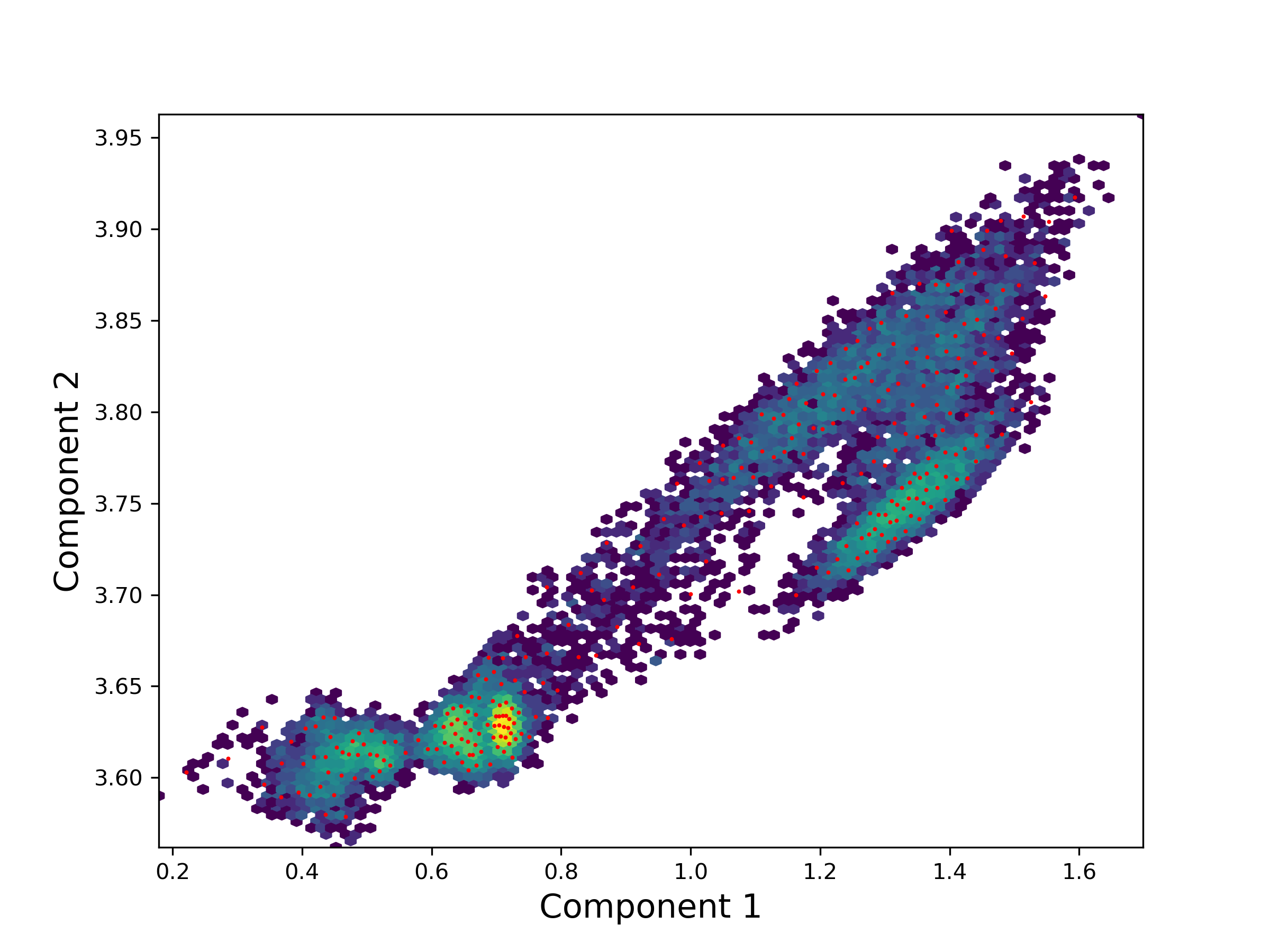}
	\caption{MiniBatch k-means clustering results of $300$ microstates shown in red dots.}
	\label{micro}
\end{figure}

Simulation trajectories were projected onto a two-dimensional (2D) plot in RMSD of C$\alpha$ atoms with reference to the close and open state structures, respectively (Figure \ref{reduced2d_timescale}A). To apply MSM analysis, MiniBatch k-means clustering was applied to divide the simulation sampling space into $300$ microstates based on the reduced-dimension plot, shown in \ref{micro}. The estimated relaxation timescale was plotted in Figure \ref{reduced2d_timescale} B. The trend of implied relaxation timescale showed that the estimated timescale is converged after $15$ ns, which was chosen as the lag time for markov state model. The number of macrostates were determined based on the band gap in estimated relaxation timescale plot.

\begin{figure*}[t]
	\begin{minipage}[c][\width]{0.49\textwidth}
	   \centering
	   \includegraphics[width=\textwidth]{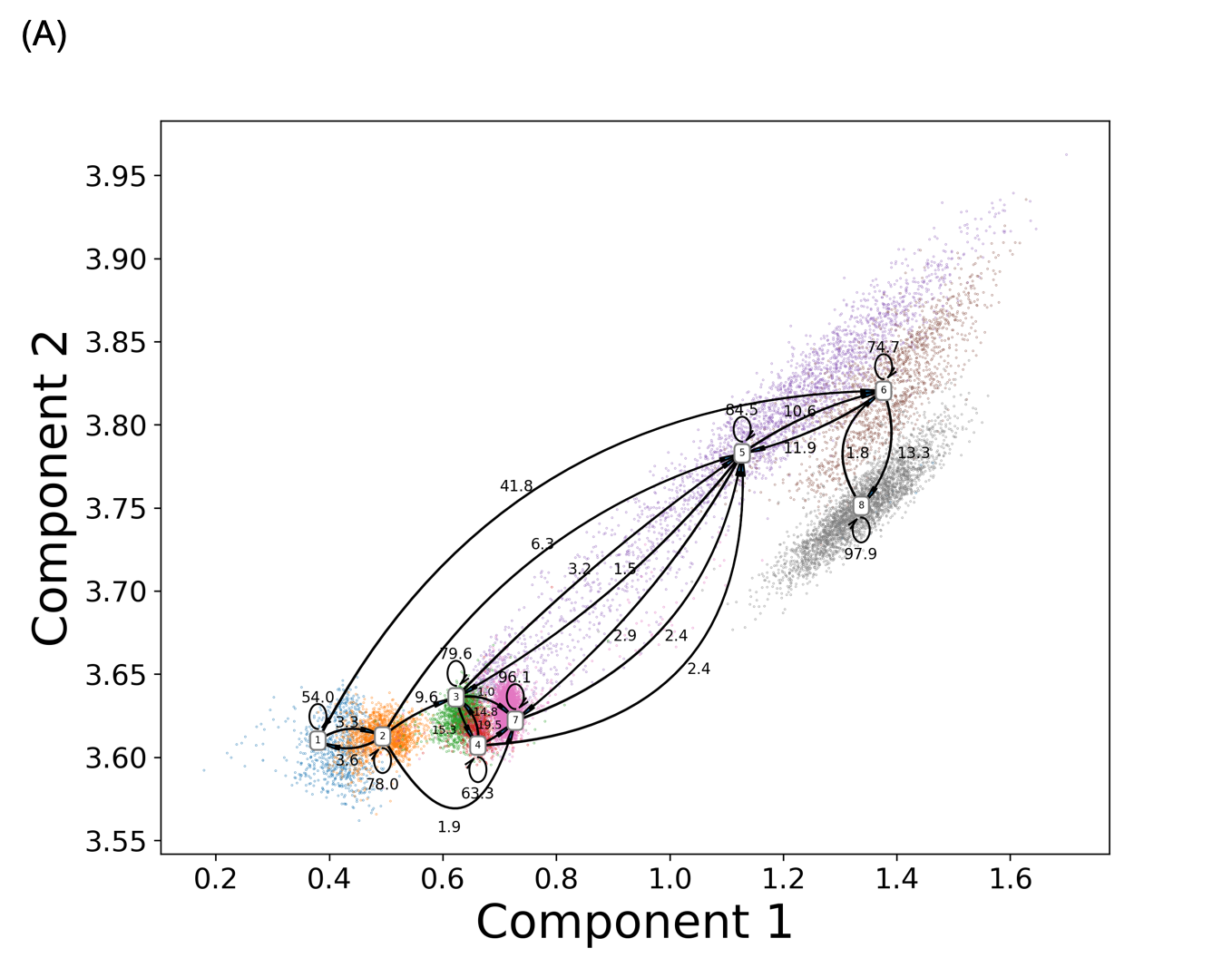}
	\end{minipage}
 	\hfill 	
	\begin{minipage}[c][\width]{0.49\textwidth}
	   \centering
	   \includegraphics[width=\textwidth]{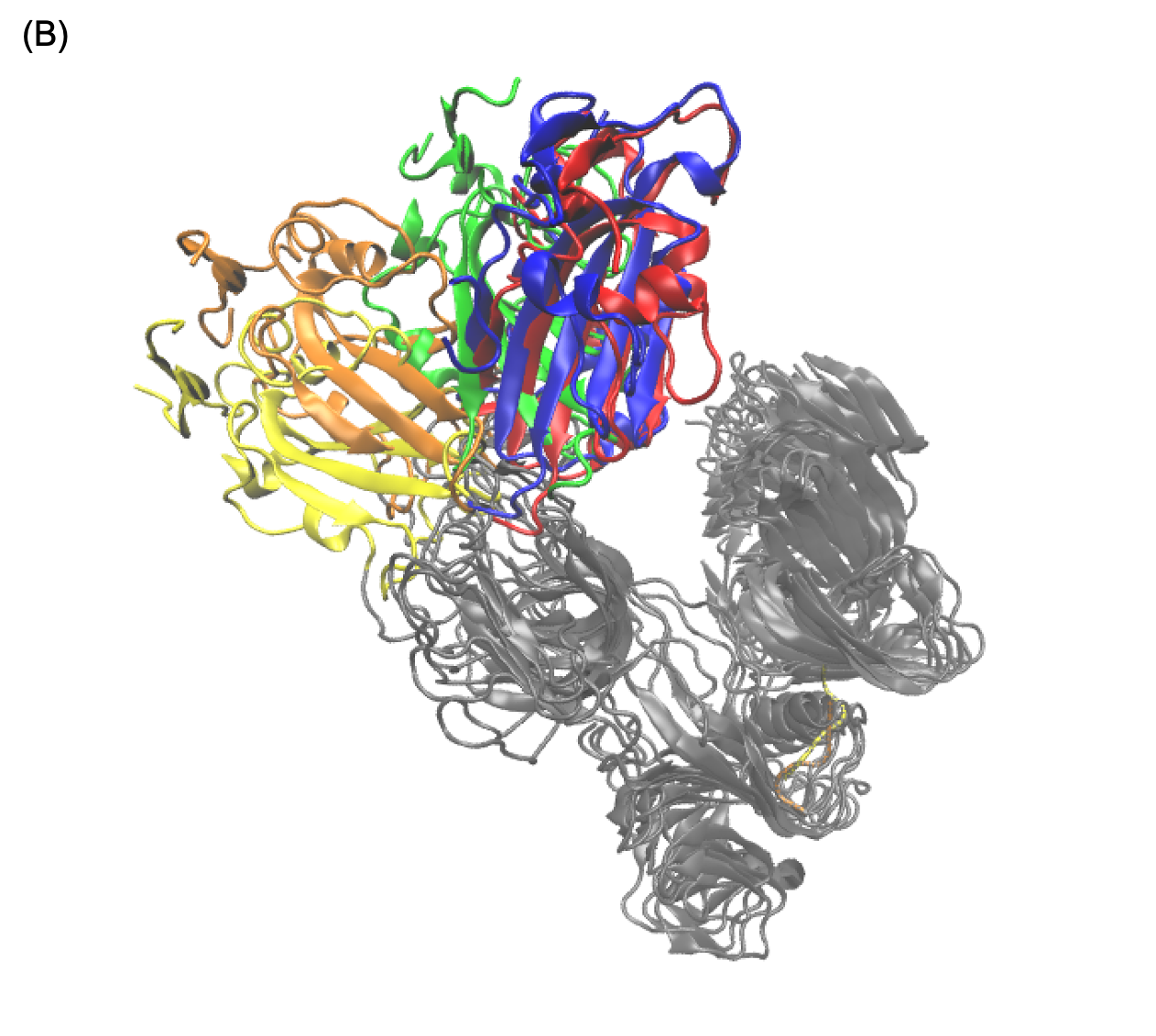}
	\end{minipage}
\vspace{-2em}
\caption{Markov state model based on the simulation: (A) Macrostates and estimated transition probabilities among them; (B) Representative structures of macrostate 2 (blue), 3 (red), 5 (green), 6 (orange) and 8 (yellow). }
\label{transition_VMD}
\end{figure*}

Total of $8$ macrostates were chosen to divide simulations into kinetically separate macrospaces. PCCA was applied to map microstates onto macrostates based on the eigenfunction structure of transition probability matrix. The resulting macrostates with transition probability are shown in Figure \ref{transition_VMD} (A). Close state and open state were equally divided into $4$ macrostates, as states 2, 3, 4, 7 belonging to the close state and states 1, 5, 6, 8 belonging to the open state. The close state is stable with $95.5\%$ probability to stay within close macrostates. \added{Macrostate $2$ was found important due to its high probabiltity of $9.9\%$ to transfer from itself to open macrostates. The average transition from closed macrostates to open macrostates is $4.5\%$.}

\begin{table*}[t]
\caption{\textbf{The probability of top 5 channels. }}
\centering
\begin{tabular}{c c}
\hline
Channels	& Probability\\
\hline
2,3,5,6,8	&23.7\%\\
2,3,4,7,5,6,8	&15.8\%\\
2,5,6,8	&11.0\%\\
2,3,7,5,6,8	&9.6\%\\
2,3,4,7,8	&8.0\%\\
Top 10 channels	&88.7\% \\
\hline
\end{tabular}
\label{channels}
\end{table*}

The first frame of the close state, which locates at State $2$ and the last frame of the partially open state, at State $8$, were chosen as the representative structures for the close and open states, respectively. Transition path theory was applied to calculate possible transition pathway connecting these two states. Total of $2,317$ pathways were generated and divided as $51$ distinct channels floating from the State $2$ to the State $8$. The probability of each channel was calculated based on net flux from initial state to the target state. Overall, the probability of top $5$ channels was listed in the Table \ref{channels}, with the contribution from the top $10$ channels accounting for $88.7\%$ of total population. The most probable path was from State $2$ $\rightarrow$ State $3$ $\rightarrow$ State $5$ $\rightarrow$ State $6$ $\rightarrow$ State $8$, and the representative structures were plotted in Figure \ref{transition_VMD}B to show a series of transition processes.

\subsection*{Random forest identifies important residues and structures}

To better understand the shift between close and open states in S1 subunit, the pair wised C$\alpha$ distances of S1 were extracted as features representing the character of protein configurations. There are $540$ residues on each chain, residue ID from $27$ to $676$, and total of $1620 * 1619 / 2 = 1,311,390$ C$\alpha$ distances were collected as features. Before further analysis, features were transformed into contact map with our proposed feature transformation technique described in Methods section. Considering the non-bonded chemical interactions length, we pick $10.0\mathring{A}$ as threshold for feature transformation.

\begin{figure*}[t]
	\begin{minipage}[c][\width]{0.49\textwidth}
	   \centering
	   \includegraphics[width=\textwidth]{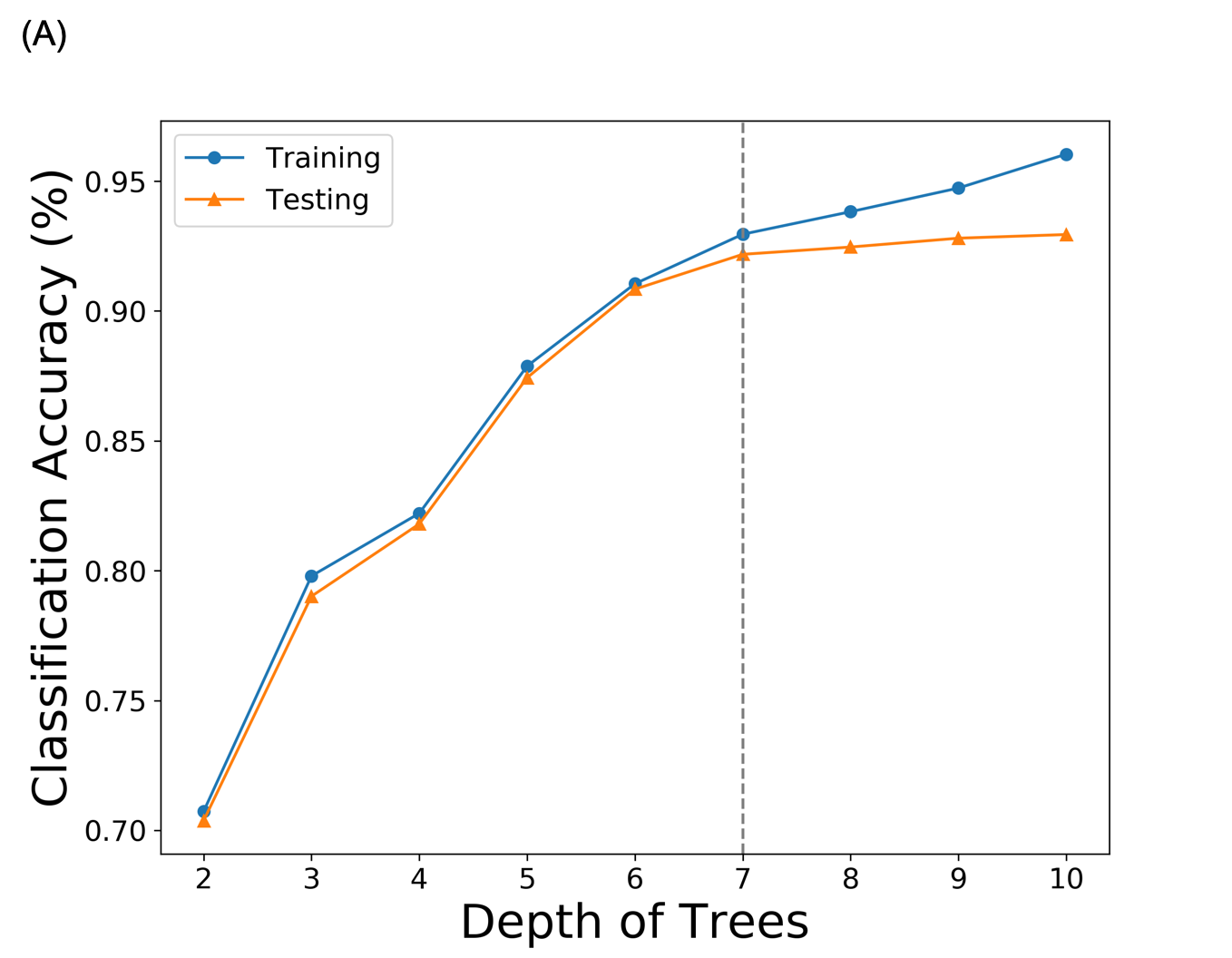}
	\end{minipage}
 	\hfill 	
	\begin{minipage}[c][\width]{0.49\textwidth}
	   \centering
	   \includegraphics[width=\textwidth]{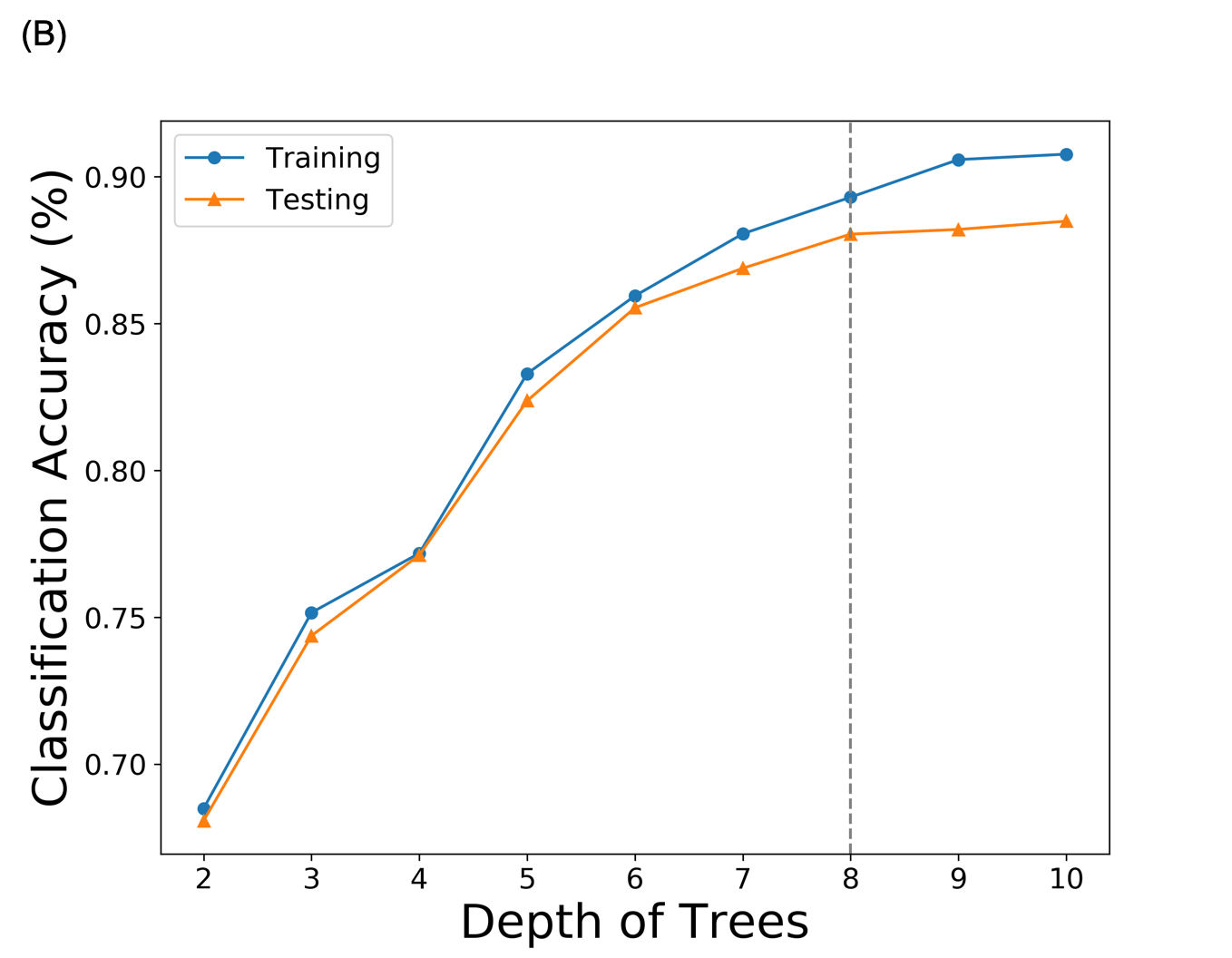}
	\end{minipage}
\vspace{-2em}
\caption{Random forest classification model using pair-wised C$\alpha$ distances in S1 subunit: (A) Classification accuracy using different depths of trees. Depth $7$ (shown in grey dashed line) was chosen with $92.18\%$ accuracy; (B) Classification accuracy regarding different depths of trees using pair-wised C$\alpha$ distances \added{within chain B and C}. Depth $8$ (shown in grey dashed line) was chosen with $88.04\%$ accuracy. }
\label{depth_rf}
\end{figure*}

\begin{figure}[ht]
	\centering
	\includegraphics[width=0.9\textwidth]{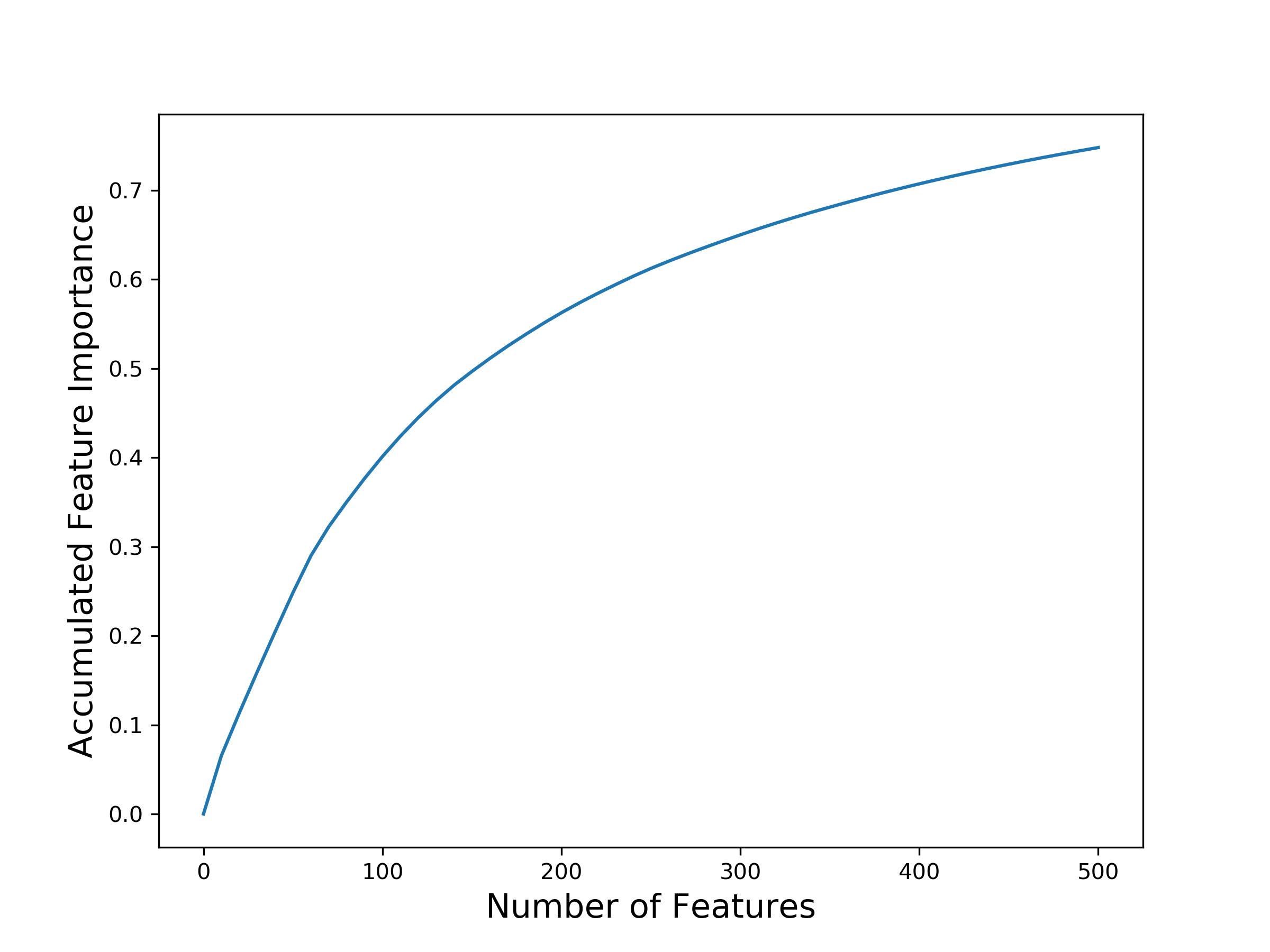}
	\caption{Accumulated feature importance of top 500 C$\alpha$ distances.}
	\label{accumulation_importance}
\end{figure}

Supervised machine learning model was applied to extract the key residues that are vital during allosteric process and study the structural differences among macrostates. For each simulation trajectory, frames were saved for every $1.2$ nanoseconds (ns), resulting in $8,334$ frames. Therefore, $16,668$ samples with $1,311,390$ features were extracted from the trajectories. Each sample was labeled based on the above macrostate result. Full dataset was further split into training set($70\%$) and testing set($30\%$). After the preparation of data, random forest model was applied to distinguish the intrinsic conformational differences among macrostates. Training scores and testing scores were plotted in Figure \ref{depth_rf}A. $7$ was chosen for the depth with corresponding testing accuracy of $92.18\%$, which indicated that the random forest model was able to catch the conformational characteristics of each macrostate only using pair-wised C$\alpha$ distances. To further investigate the relationship between chain A and two other chains, the original C$\alpha$ distances related with chain A was excluded as reduced features and applied to another random forest model. Training and testing results are shown in Figure \ref{depth_rf}B. The top $500$ features accounted for $74.8\%$ percent of the overall feature importance, shown in Figure \ref{accumulation_importance}. The testing accuracy with reduced features reached $88.04\%$ at depth $8$.

\begin{figure*}[t]
 \includegraphics[width=0.9\textwidth]{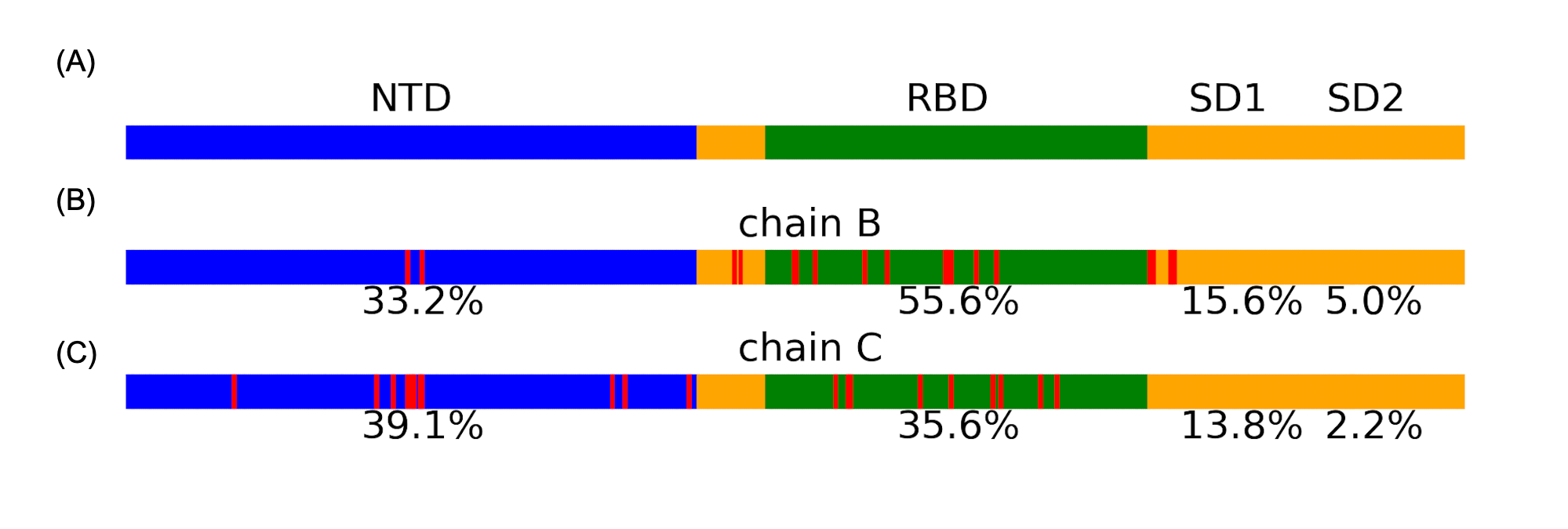}
 \caption{(A) \added{Residue sequence with tertiary structure in S1 subunit, referenced to the released structure in the prefusion conformation \cite{wrapp2020cryo}. NTD (blue), N-terminal domain; RBD (green), receptor-binding domain; SD1 and SD2 (orange), subdomains. The sequence starts with residue Ace 26 to Thr 676. (B-C) The position of top $20$ important residues are shown in red color. Accumulated tertiary structure importance in chain B and chain C are shown in numbers, respectively.} }
 \label{axis}
\end{figure*}

\begin{table*}[t]
\caption{\textbf{Top 5 C$\alpha$ distances.}}
\centering
\begin{tabular}{c c}
\hline
C$\alpha$ distances	& Importance \\
\hline
Chain C Phe 342, Chain C Asp 442 	&0.86\%\\
Chain C Ala 419, Chain C Tyr 423	&0.83\%\\
Chain B Thr 323, Chain B Thr 333	&0.76\%\\
Chain C Cys 136, Chain C Gly 142	&0.71\%\\
Chain B Leu 390, Chain B Gly 545	&0.60\%\\
\hline
\end{tabular}
\label{CAdistances}
\end{table*}

\begin{figure}[ht]
	\centering
	\includegraphics[width=0.9\textwidth]{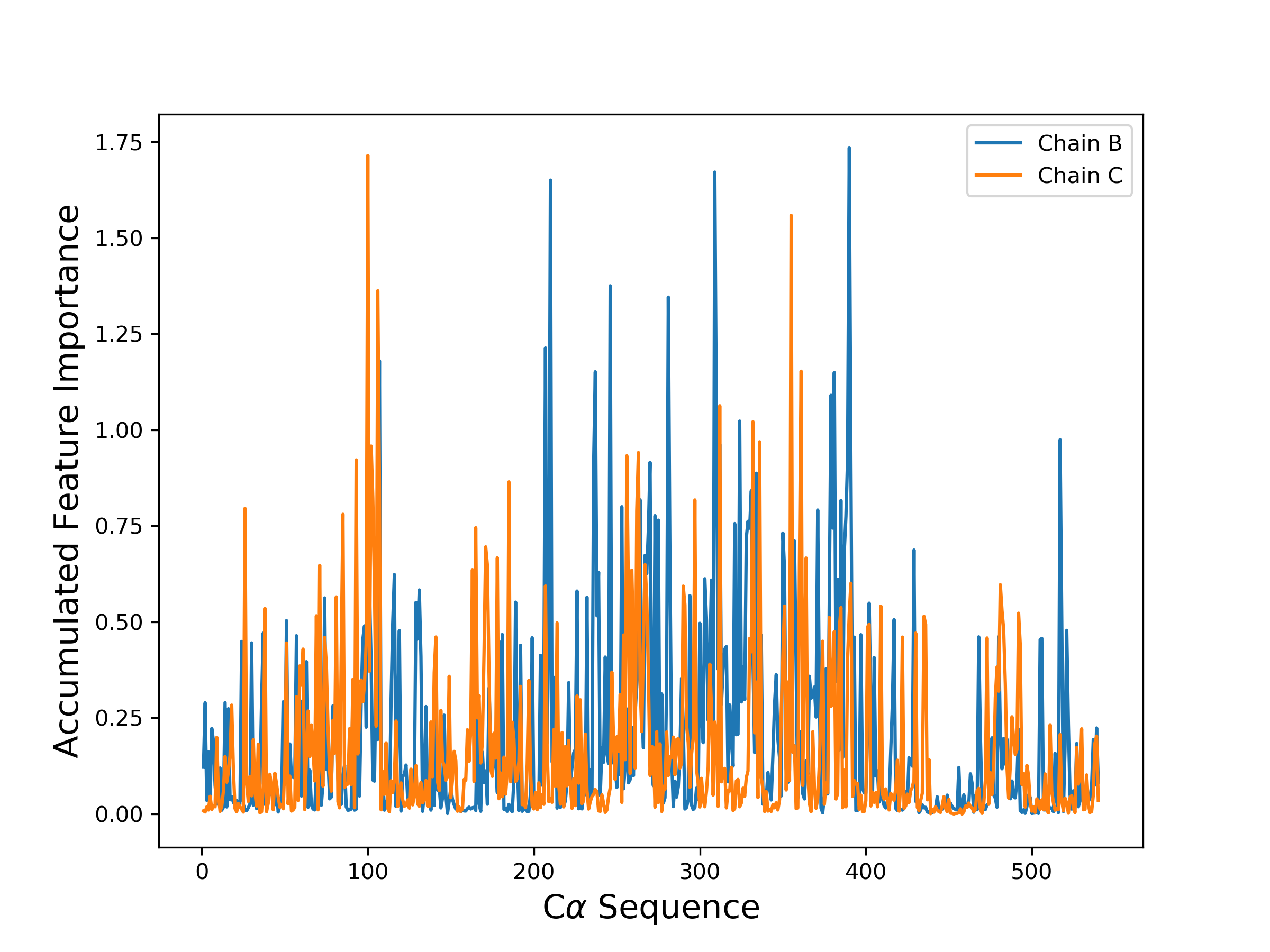}
	\caption{Accumulated residue importance in chain B (blue) and C (orange), respectively.}
	\label{chainBC}
\end{figure}

The top five important C$\alpha$ distances were listed in Table \ref{CAdistances}. In order to identify key residues along the transition from the closed to the partially open state, the feature importance of each C$\alpha$ distance was added and accumulated to the two related residues. S1 subunit structure was plotted in Figure \ref{axis}A as reference. Top $20$ important residues on chain B and C, with corresponding structure and accumulated structure importance under each figure, were plotted in Figure \ref{axis} (B-C). Full results of residue importance on chain B and C are shown in \ref{chainBC}. 

\section{Discussion}

The significance of the partially open state of receptor-binding domain (RBD) in SARS-CoV-2 for interacting with the host cell receptor has been extensively studied. \cite{walls2019unexpected,yuan2017cryo} The opening of S1 subunit, thus exposing RBD, is necessary for engaging with ACE2 and following cleavage of $S_2^{'}$ site. \cite{kirchdoerfer2018stabilized} While the RBD exhibits inherently flexibility enabling itself recognized by the receptor \cite{kirchdoerfer2016pre}, the motion of this close to open state shift still needs indepth study.

It is reported that the SARS-CoV-2 S trimer shows a $C3$ symmetry at close state and asymmetry with chain A at open state. \cite{wrapp2020cryo} Through the RMSF result, we noticed the asymmetry in dynamics at both close and open states. The S1 subunit in chain B and the S2 subunit in chain C are more dynamically active than the corresponding structures in the close state, respectively. The S1 domain in chain B is also more flexible than that in chain C in the open state. This implied the asymmetrical biological functions among the three chains.

This implication is supported by the random forest model results. Random forest model reached high accuracy in predicting macrostates based on the pair-wised C$\alpha$ distances on S1 subunit. The reason for the good performance in macrostates prediction is that the input feature contains the motion of chain A, which is sufficient to identify the main conformational change from the close to the open state. There is only a small drop of $4.14\%$ in the prediction accuracy with reduced features, which does not include the motion of chain A. Combined with the finding of asymmetric dynamics in RMSF result, we hypothesized that there is a correlation between the chain A and two other chains, causing the asymmetrical dynamics behavior. Moreover, \added{Moreover, in order to focus on the tertiary structure}, the importance of C$\alpha$ distances was accumulated to S1 domain structure and we numerically identified key structure correlated with RBD in chain B, NTD in chain C, RBD in chain B and NTD in chain B in descending order. \added{This relationship may come from the chain B and C's contribution to the protein motion in chain A. This could also originate from the protein-protein interaction along the opening movement of chain A. Further investigation of this mutual influence is warranted for a detailed clarification.}

\added{Markov state model showed a great difference in transition probability in macrostate $2$ ($78.0\%$) and $1$ ($54.0\%$), which are in closed state and partailly open state, respectively. This result implies that S1 subunit is more likely to stay in closed state, which agrees with the experimental finding that the close state is more dynamically stable than the partially open state \cite{}. However, macrostate $8$ exihibited a high probability ($97.9\%$) to stay within itself, where the RBD is detached from the S2 fusion machinery. This showed a possible dynamically stable state followed by the partially open state of the RBD and could be important in the close-open transition. Transition path theory further provided potential channels from macrostate $2$ to $8$ with the top one of $23.7\%$ probability. This channel is considered important in representing the transient shifting and can be treated as the typical protein movement. A rational design affecting the close-to-open transition can be developed through this key channel along with the important residue pairs and key tertiary structure domains.}

\section{Conclusion}

The spike protein is essential for SARS-CoV-2 as it destabilizes the trimer structure, causing the detachment of S1 subunit and exposing the RBD domain to host cell membrane. In this study, we used publicly available simulation trajectories of S protein and studied the asymmetric dynamics nature of the trimer structure. Markov state model was applied to divide the conformational space into 8 macrostates. The representative structures along the most probable channel in transition path theory are shown to present a clear route from the close state to the partially open state. Transition matrix was calculated to determine the probability of close-open transitions, with maximum of $9.9\%$ from the close macrostate $2$ to the open macrostates. The little difference between prediction accuracy results in two random forest models, where one includes the movement of chain A and the other does not, implied a relationship between chain A and two other chains. Yet, whether this relation originates from the mutual influence among chains or the intrinsic asymmetry in biological functions needs further investigation. \added{Overall, our study quantitatively analyzed the motion of S1 subunit in S protein with important C$\alpha$ distances and residues. Possible prevention and cure searching progress could be facilitated combining with these vital findings, which provide new opportunities for potential drug research and vaccine development}. 

\begin{acknowledgement}
Computational time was generously provided by Southern Methodist University's Center for Scientific Computation. The authors thank D. E. Shaw for sharing the SARS-CoV-2 spike glycoprotein trajectories. 
\end{acknowledgement}

\newpage
\bibliography{ref.bib}

\newpage
\beginsupplement
\section*{Supporting Information Available}

\end{document}